\documentclass{emulateapj}
\usepackage{lscape}

\shorttitle{Imaging of secondary components}
\shortauthors{Tokovinin}
\begin{document}

\renewcommand{\topfraction}{1.0}
\renewcommand{\bottomfraction}{1.0}
\renewcommand{\textfraction}{0.0}

\title{Imaging    survey   of  subsystems in  secondary   components    to nearby
  southern dwarfs\altaffilmark{\dag}}

\altaffiltext{\dag}{Based  on observations  obtained  at the  Southern
  Astrophysical Research (SOAR) telescope, which is a joint project of
  the Minist\'{e}rio da Ci\^{e}ncia, Tecnologia, e Inova\c{c}\~{a}o da
  Rep\'{u}blica  Federativa  do  Brasil,  the  U.S.  National  Optical
  Astronomy Observatory,  the University  of North Carolina  at Chapel
  Hill, and Michigan State University. }

\author{Andrei Tokovinin}
\affil{Cerro Tololo Inter-American Observatory, Casilla 603, La Serena, Chile}
\email{atokovinin@ctio.noao.edu}

\begin{abstract}
To  improve  the statistics  of  hierarchical multiplicity,  secondary
components  of wide  nearby  binaries with  solar-type primaries  were
surveyed  at  the  SOAR  telescope  for evaluating  the  frequency  of
subsystems.   Images of 17  faint secondaries  were obtained  with the
SOAR Adaptive Module that improved the seeing; one new 0\farcs2 binary
was detected. For all targets, photometry in the $g', i', z'$ bands is
given.    Another   46    secondaries   were   observed   by   speckle
interferometry, resolving 7 close subsystems.  Adding literature data,
the binarity of  95 secondary components is evaluated.   We found that
the detection-corrected frequency of secondary subsystems with periods
in the well-surveyed range from $10^3$ to $10^5$ days is 0.21$\pm$0.06
-- same  as the  normal frequency  of such  binaries  among solar-type
stars, 0.18.   This indicates  that wide binaries  are unlikely  to be
produced by  dynamical evolution of  $N$-body systems, but  are rather
formed by fragmentation.
\end{abstract}


\section{Introduction}
\label{sec:intro}

This   paper  complements   multiplicity  statistics   in   the  solar
neighborhood.   Recently, multiplicity  data  on the  F- and  G-dwarfs
within  67\,pc of  the Sun  (the FG-67  sample) and  their statistical
analysis     were    published     \citep[][hereafter     FG67a    and
  FG67b]{Tok2014a,Tok2014b}.   This work revealed  that the  census of
subsystems in  the {\it secondary} components of  nearby wide binaries
is much less complete than for the main (primary) targets.  To address
this  problem, a  large  survey  of 212  secondary  components on  the
northern  sky  has been  undertaken  with  the  Robo-AO instrument  at
Palomar \citep{RoboAO}.  On the  southern sky, however, only a limited
one-night survey  of wide binaries  was made with the  NICI instrument
\citep{THH10}  and a  few secondaries  were addressed  individually by
various authors.

Here  we  imaged 17  faint  secondary  components  at the  4.1-m  SOAR
telescope using  the laser-assisted  adaptive optics (AO) system  SAM (SOAR
Adaptive  Module) \citep{SAM10,SAM12}  to  improve spatial  resolution
with respect to the seeing.  Although the achieved resolution of about
0\farcs5 is  inferior to the  0\farcs1 resolution of the  Robo-AO, the
wide corrected field of SAM allows comparison of the target image with
other stars  and enables  binary detection down  to a fraction  of the
Full-Width      at       Half      Maximum      (FWHM)      resolution
\citep[see][]{Terziev2013}.   The  secondaries  probed  with  SAM  are
fainter  than those  surveyed  with Robo-AO,  extending the  subsystem
census into the low-mass regime.

In  addition to  the  AO-assisted classical  imaging,  we targeted  46
brighter  secondary components  with  the speckle  camera. Here,  SOAR
reaches  the diffraction-limited  resolution of  0\farcs04 in  the $I$
band, surpassing  Robo-AO.  Thus, this survey  complements the Robo-AO
effort  in different  ways, while  extending it  to the  southern sky.
Joining new observations with data  from the literature, we cover here
95   secondary  components  and   give  a   statistically  independent
assessment of the frequency of the secondary subsystems.

The fraction of subsystems in  the secondary components is a sensitive
probe  of formation mechanisms  of binary  stars. Chaotic  dynamics of
small  $N$-body systems  leads  to preferential  ejection of  low-mass
stars.  Some  ejected stars remain  bound to their  massive primaries,
but  they  tend  to  be   single  and  their  wide  orbits  have  high
eccentricity  \citep[see  however  Fig.~3 in  ][]{Reipurth2012}.   The
alternative   formation   mechanism    of   multiple   stars   through
fragmentation of  rotating cores  predicts that components  on distant
orbits inherit a  large fraction of the total  angular momentum of the
core. Therefore their orbits should have moderate eccentricities, they
never come  very close  to the main  star, and no  dynamical interplay
takes place.  In such case,  the secondary components are as likely to
contain subsystems as the  primaries. This conclusion emerges from the
study of the FG-67 sample; it is strengthened here by the new data.

In Section~\ref{sec:SAM} we  present observations of faint secondaries
with  SAM.    The  results  of   the  speckle  survey  are   given  in
Section~\ref{sec:speckle}. The complete sample of secondaries is given
in  Section~\ref{sec:data}, its statistical  analysis is  described in
Section~\ref{sec:stat}. Section~\ref{sec:disc} discusses the results.

\section{Observations with SAM}
\label{sec:SAM}

\subsection{Observing procedure}
\label{sec:samobs}

The  night of  March 4,  2014 was  allocated for  the survey  with SAM
through NOAO (proposal 2014A-0039).  We selected from the FG-67 sample
physical   secondary   components   fainter  than   $V=15$\,mag   with
separations larger than 20\arcsec~ and no prior high-resolution data.

\begin{figure}
\epsscale{1.0}
\plotone{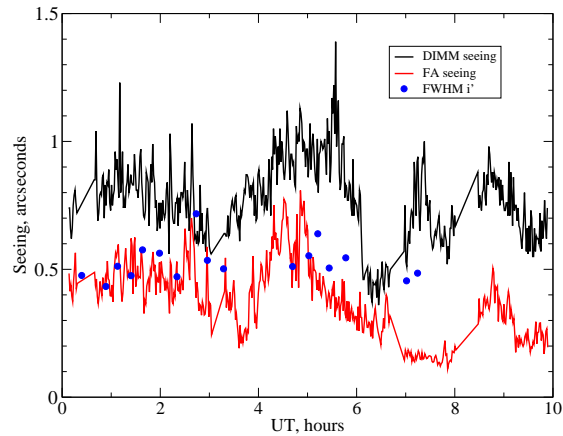}
\vspace*{0.5cm}
\caption{Observing conditions  on March 4/5,  2014. The black  and red
  lines  show  the total  seeing  and  the  free-atmosphere seeing  as
  measured, respectively,  by the DIMM  and MASS channels of  the site
  monitor  at Cerro  Pach\'on. The  dots show  FWHM resolution  of the
  images in closed loop in the SDSS $i'$ filter.
\label{fig:seeing} 
}
\end{figure}

The observing conditions on March 4/5, 2014 were typical for the Cerro
Pach\'on site (Figure~\ref{fig:seeing}).  The seeing fluctuated around
its median value of 0\farcs7,  while the seeing in the free atmosphere
(produced  by  turbulence above  0.5\,km)  was  mostly below  0\farcs5
\citep[its  median at  Cerro Pach\'on  is  0\farcs40,][]{TT08}.  There
were light cirrus  clouds at the beginning of  the night which however
did not prevent  laser operation.  The observations had  to be stopped
at UT  7:30 when the laser  projection optics was damaged  by a burned
insect. In  the following runs we  installed a protective  mesh in the
laser  launch telescope  to  prevent such  incidents, without  adverse
effect on the SAM performance.  A total of 21 targets (17 of those for
this  program) were  pointed, with  a median  overhead time  (from the
start of the telescope slew to  closing all loops) of 7\,min.  The SAM
operated  thus  quite efficiently.   It  delivered  a FWHM  resolution
limited      mostly       by      the      free-atmosphere      seeing
(Figure~\ref{fig:seeing}).

\subsection{Data reduction}
\label{sec:samdat}

The images were acquired by  the CCD of 4096$\times$4112 pixels binned
2$\times$2 to the effective pixel scale of 91\,mas. It covers a square
field of 186\arcsec. For each target, three images in the SDSS filters
$g'$,  $i'$, and  $z'$ \citep{Fukugita1996}  were taken  with exposure
times  ranging from 10\,s  to 1\,min.   The data  were processed  in a
standard way  (bias subtraction and  division by sky flats)  using the
pipeline PyRAF  script of L.~Fraga.  Then three images in  each filter
were median-combined.

\begin{figure}
\epsscale{1.0}
\plotone{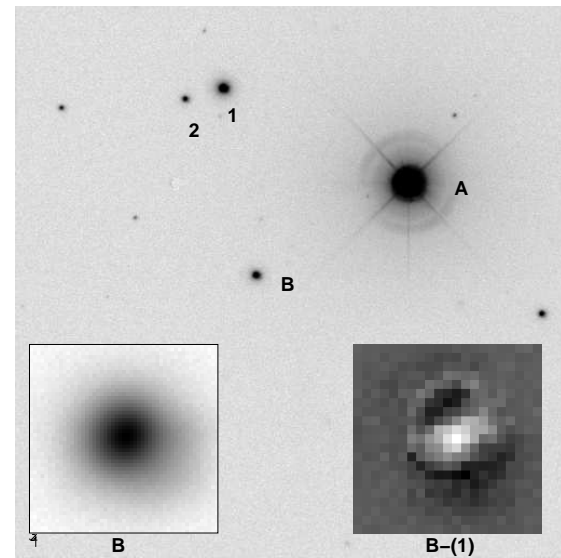}
\caption{Example of the  SAM data and their processing.   The image of
  HIP~50895  in the  $i'$  filter, in  negative logarithmic  intensity
  scale (North up, East to the right) is shown.  The binary components
  A (heavily saturated) and B are marked, as well as the two reference
  stars (1) and  (2).  The lower-left insert shows  the enlarged image
  of  the  component B  (also  in  logarithmic  intensity scale),  the
  lower-right  insert  shows   the  residuals  after  subtracting  the
  reference star (1) from the image of B (sub-frame of 21 pixels size,
  linear gray scale from $-0.04$ to $+0.073$ of the PSF maximum).
\label{fig:sam} 
}
\end{figure}

Figure~\ref{fig:sam}  illustrates  typical   data.   The  wide  binary
HIP~50895AB  with   a  separation  of  42\farcs4   was  identified  by
\citet{LEP}. The components share  common proper motion and the colors
of B match a  dwarf of 0.2 solar mass located at  the same distance as
A,  54\,pc.   The  $V$ magnitudes  of  A  and  B  are 8.12  and  16.3,
respectively, so the image of A  is heavily saturated.  The sky is not
crowded, but  several field stars  are still available as  point
spread  function (PSF)  reference.  For  each secondary  component, we
selected at least two reference stars.

\begin{figure}
\epsscale{1.0}
\plotone{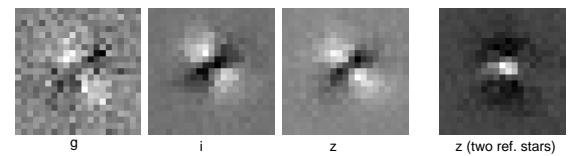}
\caption{Detection of the close binary HIP~53172 Ba,Bb.  The images display
  residuals from  fitting the  target with the  reference star  (1) in
  three bands.  The last  image shows residuals from fitting reference
  star (1) with reference star (2).
\label{fig:HIP52172} 
}
\end{figure}

\subsection{Detection of binary companions}
\label{sec:sambin}

We  developed  several  tools  to detect  and  characterize  potential
subsystems (custom software in IDL).  First, all targets and reference
stars  were fitted  by the  Moffat function.   The FWHM  resolution was
determined from these fits, while  residuals show any asymmetry of the
PSF.  The  PSFs have a very  faint ``tail'' on  their lower-right side
related to the deformable mirror in SAM.  Otherwise, the PSFs are very
symmetric,  with ellipticity well  under 0.1.  The peak  intensity and
total flux were determined for  all targets and reference stars in all
three filters.

Each target was fitted by a  scaled and shifted image of the reference
star. This is the most  sensitive test for binary companions. The fits
were repeated  with the second  reference star, and the  two reference
stars were mutually  cross-checked as well. We restricted  the fits to
the 10-pixel  radius, looking  for close companions  (wider companions
are evident anyway). The quality of  the fit could be evaluated by the
normalized  $\chi^2$ metric  if the  residuals were  dominated  by the
readout  and  shot noise.   However,  the  major  contribution to  the
residuals  comes from  slight  differences between  the  PSFs, so  the
adequate  metric of  the fit  quality is  the rms  residual difference
normalized by the  intensity. If $A_i$ and $B_i$ are
intensities of the target and fitted PSF, respectively, in the subset
of ${i}$ pixels, the fit quality metric $r$ is
\begin{equation}
r^2 = \sum_i (A_i - B_i)^2 / \sum_i A_i^2 . 
\label{eq:r}
\end{equation}
The median value of $r$ in all  filters is about 0.05, and it does not
exceed  0.1.   The  lower-right  insert in  Figure~\ref{fig:sam}  shows
the residual  pattern with  $r=0.047$: a  bright central zone and  a dark
halo.  The AO correction was  slightly better for the target star than
for  the  reference (1)  (FWHMs  of  0\farcs72 and  0\farcs75,
respectively), causing the residual mismatch. A similar pattern is seen
when the  target is  fitted by the  reference (2), $r=0.051$.   On the
other hand, the  two reference stars (1) and (2)  are located close to
each other and match better, $r=0.021$. 

Among the 17 observed targets,  we detected only one new subsystem in
HIP~53172B.  This is a late-M  dwarf with estimated mass of $0.1 {\cal
  M}_\odot$  located   at  279\arcsec~  from  the   main  component  A
\citep{LEP}. Its distance from the  Sun is 47\,pc, $V$-magnitudes of A
and B are  7.76 and 19.8, respectively (A was  outside the SAM field).
The  residuals to  the PSF  fits in  all filters  consistently  show a
``butterfly''   pattern  expected   for   an  equal-component   binary
(Figure~\ref{fig:HIP52172}), while the  two reference stars match each
other  better.  The  same  asymmetry is  seen  when approximating  the
target  by  a  Moffat  function.   This detection  is  not  absolutely
certain, but very likely.

The parameters of the binary pair HIP~53172 Ba,Bb were determined by a
procedure similar to the PSF  fit.  Instead of adjusting only relative
position and intensity ratio to  the PSF star, the fitting routine now
assumes  that the  target is  a  binary and  adjusts three  additional
parameters  (relative  position of  the  binary  components and  their
intensity  ratio).  In  the  case  of HIP~53172  Ba,Bb,  we found  the
separation of $\sim$2 pixels (0\farcs18)  at a position angle of $\sim
320^\circ$, with equal intensity  ($\Delta m = 0$).  Consistent binary
parameters are  found in all three  filters and while  using either of
the two reference stars.  The FWHM resolution is 0\farcs5 in both $i'$
and $z'$ filters.  So, the binary  separation is less than half of the
FWHM, its  measurement is  quite uncertain and  cannot be used  in the
future  orbit  calculation (the  estimated  orbital  period  is a  few
decades).   The quality  of  the PSF  fits  in the  $z'$ filter  using
reference  stars (1)  and (2)  is 0.063  and 0.095,  respectively.  By
fitting  the double-star  model, the  residuals improve  to  0.033 and
0.027.  Similar improvement  of residuals is found in  the $i'$ filter
when fitting a double star instead of the PSF.

\begin{figure}
\epsscale{1.0}
\plotone{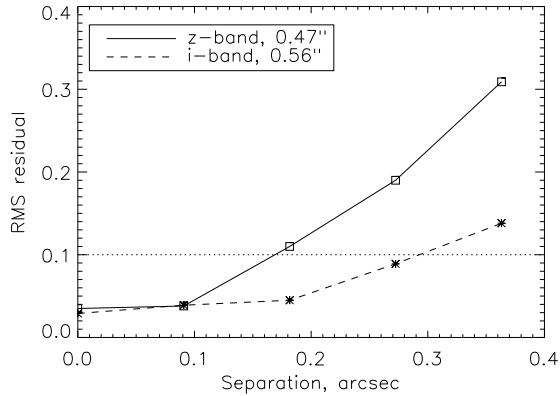}
\caption{Sensitivity  of the  rms residual  $r$ from  the PSF  fit (on
  vertical axis) to binary  separation. The target HIP~43279B was used
  to  simulate binaries  with equal  components  and fit  them by  the
  reference star (1). The legend gives the FWHM resolution.
\label{fig:rms-sep} 
}
\end{figure}

To determine the sensitivity  to subsystems, we simulated binaries and
fitted them  by reference stars. By adopting  a conservative detection
threshold  of $r>0.1$, we  found that  an equal-magnitude  binarity is
securely detected at a separation equal to half of the FWHM resolution
(Figure~\ref{fig:rms-sep}), while  at a  separation equal to  the FWHM
the detectable intensity ratio is about  0.16, or $\Delta m < 1.8$. We
further  adopt $\Delta m  < 3.2$  at 0\farcs7  and $\Delta  m <  5$ at
5\arcsec, independently of the FWHM. The deepest detection in terms of
the  mass ratio  is in  the band  $z'$, where  the FWHM  resolution is
better  and the  low-mass  companions are  brighter.  These  detection
limits are adopted in  the statistical analysis presented below.  They
are  obviously approximate  and conservative  (the new  pair HIP~53172
Ba,Bb  is just  below  the limit).   To  verify the  absence of  other
detectable close  companions, we  ran the binary-fitting  algorithm on
all targets.  It returned ``binaries'' with separations of less than 2
pixels. There was no  agreement between binary parameters in different
filters,  while  the  residuals  from  fitting double  stars  were  not
substantially reduced in comparison to the PSF fits.

\subsection{Photometry}
\label{sec:samptm}

For  each  field,  we  determined  instrumental  magnitudes  of  stars
detected in all three filters by aperture photometry with the aperture
radius of 10 pixels and the sky radius of 20 pixels. The zero point of
the  instrumental magnitudes  corresponds to  25~mag for  a flux  of 1
count per second.

Four targets in the equatorial zone covered by the SDSS (HIP~43172,
43297, 56738,  60081) were used for photometric  calibration. Stars in
these fields were  matched to the SDSS Data Release  9 \citep{DR9} using the
TOPCAT                                                  tool.\footnote{
  \url{http://www.star.bris.ac.uk/\~{}mbt/topcat/}}    We    calibrate
instrumental magnitudes against the  PSF magnitudes of SDSS.
After rejecting  a few outliers and  stars fainter than $g'  = 19$, we
got  22  matches and  fitted  the  instrumental  magnitudes by  linear
relations like
\begin{equation}
g'_{\rm inst}  = g'_{\rm SDSS} + a + b \; (g' - i' )_{\rm SDSS} .
\label{eq:ptm}
\end{equation}
The $(g'  - i' )_{\rm  SDSS}$ color term  is used for all  filters.  The
extinction  was  not  considered  explicitly, being  included  in  the
zero-points $a$, because the range of air mass was small (from 1.04 to
1.27) and  we did not measure  the extinction. Table~\ref{tab:photsys}
gives the zero  points and color terms (of  which the only significant
one  is in  $i'$) and  the rms  residuals to  those linear  fits.  The
linear  equations were inverted  to translate  instrumental magnitudes
into the standard SDSS system.  Photometry is not the main goal of our
program, just a by-product,  and its accuracy is about $\pm$0.05\,mag.

\begin{deluxetable}{ c  ccc    }           
\tabletypesize{\scriptsize}         
\tablecaption{Reduction to the SDSS photometric system
\label{tab:photsys} }                    
\tablewidth{0pt}  
\tablehead{
 \colhead{Filter} &  
\colhead{$a$} &
\colhead{$b$}  &
\colhead{rms}  
}
\startdata 
$g'$ & $-0.609\pm 0.025$ & $0.008\pm0.018$      & 0.066 \\ 
$i'$ & $-0.292\pm0.013  $ & $-0.054\pm 0.08 $    & 0.029 \\ 
$z'$ & $0.172\pm 0.019 $  & $0.009\pm 0.014 $     & 0.052    
\enddata
\end{deluxetable}

\begin{deluxetable}{ l  ccc cc   }           
\tabletypesize{\scriptsize}         
\tablecaption{Photometry of secondary components with SAM
\label{tab:sam} }                    
\tablewidth{0pt}  
\tablehead{
 \colhead{Name} &  
\colhead{$g'$} &
\colhead{$i'$}  &
\colhead{$z'$} & 
\colhead{Air} &
\colhead {FWHM} \\
  &
(mag) & (mag) & (mag) & 
mass  &
(arcsec)
}
\startdata 
28267D  &  16.90 &  13.58 &  12.76 &   1.12 &  0.58 \\
32650C  &  17.33 &  14.39 &  13.76 &   1.27 &  0.41 \\
41211B  &  16.96 &  13.73 &  12.92 &   1.15 &  0.47 \\
41211C  &  18.08 &  14.32 &  13.22 &   1.17 &  0.40 \\
43172D  &  18.98 &  15.64 &  14.72 &   1.20 &  0.50 \\
        &{\it 18.91} &  {\it 15.62} & {\it 14.70} & &  \\  
43297B  &  15.93 &  13.00 &  12.16 &   1.23 &  0.42 \\
        &{\it 16.00} &{\it 16.03?} &{\it12.21} &  & \\
49520C  &  17.74 &  14.83 &  14.09 &   1.11 &  0.52 \\
50895B  &  17.04 &  14.18 &  13.46 &   1.13 &  0.66 \\
53172B  &  18.88 &  15.62 &  14.77 &   1.13 &  0.54 \\
54530B  &  17.62 &  17.09 &  17.17 &   1.09 &  0.44 \\
55455B  &  16.72 &  13.62 &  12.90 &   1.26 &  0.59 \\
56738B  &  17.03 &  13.79 &  12.92 &   1.15 &  0.42 \\
        &{\it 17.04}&{\it  13.84} &{\it 12.90} &  & \\
57443B  &  13.87 &  10.87 &  10.08 &   1.04 &  0.53 \\
60081B  &  17.69 &  17.13 &  17.36 &   1.15 &  0.47 \\
        &{\it 17.69}&{\it  17.29}&{\it  17.37} &  & \\ 
60620C  &  16.63 &  13.97 &  13.29 &   1.06 &  0.47 \\
64056B  &  16.29 &  13.59 &  12.91 &   1.04 &  0.42 \\
66530B  &  15.71 &  13.03 &  12.39 &   1.01 &  0.43 \\
\enddata
\tablecomments{The SDSS-DR9 photometry is given in italics.}
\end{deluxetable}

Table~\ref{tab:sam}  lists 17 secondary  components observed  with SAM
(their  equatorial  coordinates  are given  in  Table~\ref{tab:list}),
their measured SDSS magnitudes $g'$, $i'$, and $z'$, the air mass, and
the FWHM resolution  in the $z'$ band.  The  SDSS magnitudes available
for  the   4  targets   in  common  with   this  work  are   given  in
Table~\ref{tab:list}  in italics;  they  agree well,  except the  $i'$
magnitude of HIP~43279B which appears to be corrupted in the SDSS.  On
the other hand, the $V$ magnitudes  estimated from $g'$ as $V_g = g' +
0.297 - 0.366(g'  - Ks) + 0.025(g' -  Ks)^2$ differ substantially from
the $V$ magnitudes listed in FG67a; the listed magnitudes are based on
the  photographic  photometry of  \citep[][]{LEP}.   For the  resolved
binary  HIP~53172,  we estimate  $V_g=17.86$  instead  of 19.8  quoted
above.

\begin{figure}
\epsscale{1.0} 
\plotone{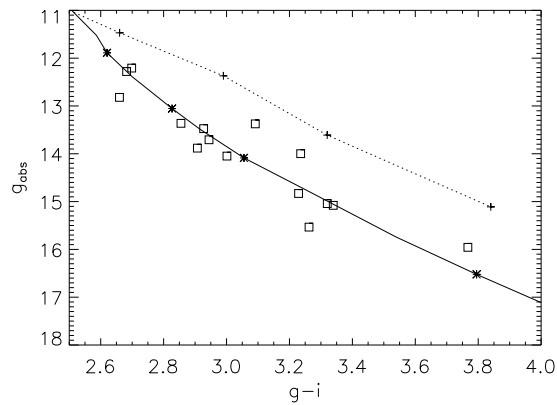}
\caption{Color-magnitude diagram of  secondary components (squares) in
  the  SDSS $g'$  and  $i'$  bands.  The  full  line shows  polynomial
  relations from FG67a with  wavelengths of 520\,nm and 770\,nm, where
  asterisks mark masses  of 0.3, 0.2, 0.15, and  0.1 ${\cal M}_\odot$.
  The  dashed  line is  based  on  the  Table~3 of  \citet{Covey2007},
  pluses mark spectral types from M3V to M6V.
\label{fig:cmd} 
}
\end{figure}

Our  program included  two  white dwarf  secondaries  (HIP 54530B  and
60081B)  which  differ from  the  remaining  stars  by their  ``blue''
colors.  We did  not find any close companions  to those white dwarfs.
The statistical analysis  in Section~\ref{sec:stat} considers only the
remaining 15 red-dwarf secondaries.

Figure~\ref{fig:cmd}  presents the  color-magnitude  diagram (CMD)  of
low-mass secondary components, constructed using the known distance to
their primary  components. For  low-mass stars, standard  relations in
the  SDSS colors are  not well  established.  The  dashed line  is the
polynomial relation from FG67a where the effective wavelengths of $g'$
and  $i'$ are  chosen  to  be 520\,nm  and  770\,nm, respectively,  to
roughly match  the data.   The main sequence  based on the  Table~3 of
\citet{Covey2007}  is  plotted  in  dashed line.   The  luminosity  of
low-mass stars depends on their metallicity (which is not measured for
most primary targets) and age as well as on mass, therefore the points
in    Figure~\ref{fig:cmd}    do   not    align    along   a    single
sequence. Equal-mass  binaries are readily detected  by their position
in   the   CMDs   of   open   clusters   \citep[see   e.g.    Figure~4
  of][]{Sarro2012},  but this  method  does not  work  for this  field
sample.

\section{Speckle observations}
\label{sec:speckle}

In January--March  2014, speckle interferometry was  performed at SOAR
to follow  the orbital motion  of close and ``fast''  visual binaries.
During  five  allocated  nights,  we  occasionally  pointed  secondary
components in  wide nearby binaries  to complement the work  done with
SAM. Although the  speckle camera was mounted on  SAM, the laser guide
star was not used in order  to maintain high efficiency (150 stars per
night on  average).  As  the secondary components  are red  and mostly
faint, they  were observed in the $I$  filter with a field  of view of
3\arcsec.  A total of  46 secondary components with separations larger
than 3\arcsec~  were pointed.  These observations had  a low priority,
being a ``filler'' in the main speckle program.

The speckle camera and  data reduction are described by \citet{TMH10}.
The  speckle  data  pipeline  evaluates maximum  detectable  magnitude
difference $\Delta  m$ at separations  of 0\farcs15 and  1\arcsec.  At
the diffraction limit (0\farcs042  at 770\,nm), a detectable binary is
assumed to have $\Delta m < 0.5$. Among the 46 observed secondaries, 7
were resolved for the first  time, while three more contain previously
known  resolved subsystems.   The measurements  of  resolved secondary
components will  be published later  with the rest of  the binary-star
measurements. Here  we give only the relevant  information, namely the
detection  limits  for  all  observed  secondary  components  and  the
separation and magnitude difference  of the newly resolved subsystems.
Most of  the new pairs were  confirmed on other dates  or in the
$y$ filter;  some  of them were already known  as spectroscopic and/or
acceleration binaries.

\section{Combined data}
\label{sec:data}

For  the statistical  analysis,  we combine  the secondary  components
surveyed here with data from the literature. We selected from the main
FG-67 database secondary  components for which high-resolution imaging
is available,  south of  $+15^\circ$ declination, and  with separations
larger  than  3\arcsec~  from  the primary  targets.   The  components
surveyed by Robo-AO were  excluded to make this analysis statistically
independent.   Apart from  this work,  the  next largest  data set  is
furnished by the mini-survey with NICI \citep{THH10}.

Table~\ref{tab:list}  lists  the   complete  sample  of  95  secondary
components  discussed  here.   Its  first  columns  contain  the  {\it
  Hipparcos}  number   of  the  main  FG-67   target  HIP1,  component
identification,  its   separation  from  the  main   target,  and  the
approximate equatorial coordinates in degrees as given in FG67a.  Then
follow the wavelength of imaging  data in nm and the detection limits
(4 separations and 4 values  of $\Delta m$).  The last column contains
the reference code, explained in the notes to the table.  The code SAM
refers  to  the  data  of Section~\ref{sec:SAM},  SOAR  means  speckle
observations (Section~\ref{sec:speckle}).

The  data on  secondary subsystems  in  this sample  are collected  in
Table~\ref{tab:sec}.   Its  first   column  identifies  the  secondary
component.  The  second column gives  the discovery method(s)  ('a' --
astrometric  acceleration,  's,S' --  spectroscopic,  'v,V' --  direct
resolution). For resolved  subsystems, we give in the  columns (3) and
(4)  angular separation  $\rho$  and magnitude  difference $\Delta  m$
together with the filter to  which it refers.  The orbital periods and
component's masses are estimated  as explained in FG67a.  Comments and
references are  provided in the  last column and  in the notes  to the
table. Known visual pairs are identified by their ``discoverer codes''
given  in the  WDS \citep{WDS}.   For these  pairs, the  WDS detection
limits in Table~\ref{tab:sec} are  adopted from FG67a, e.g.  $\Delta m
= 2.5$ at 0\farcs15.

\begin{deluxetable*}{lcr  rr c cccc cccc ll }  
\tabletypesize{\scriptsize}         
\tablecaption{List of secondary components with high-resolution imaging data (fragment)
\label{tab:list} }  
\tablehead{
\colhead{HIP1} & &
\colhead{Sep.} &
\colhead{RA(2000)} &
\colhead{Dec(2000)} &
\colhead{$\lambda$} &
\multicolumn{4}{c}{$\rho_1$ to $\rho_4$ } &
\multicolumn{4}{c}{$\Delta m_1$ to $\Delta m_4$} & 
\colhead{Ref.} \\
&    &   
(arcsec) &
(deg) & (deg) &
(nm) & 
\multicolumn{4}{c}{(arcsec)} & 
\multicolumn{4}{c}{(mag)} & 
}
\startdata
  9902& B&   52.2&   31.8843& $-$59.6726& 1690&   0.118&   0.306&   1.969&   6.0 &   0.0&   3.9&   7.9&   7.9& Cvn2010\\
 10579& B&    6.7&   34.0369& $-$21.0076& 2272&   0.054&   0.140&   0.900&   9.0 &   0.0&   3.9&   7.9&   7.9& NICI\\
 14954& B&    6.8&   48.1925& $-$1.1977& 2200&   0.060&   1.000&   2.000&   5.0 &   1.0&   6.3&   8.6&   9.9&  Mug2009\\
 15371& B&  309.1&   49.4423& $-$62.5753&  540&   0.042&   0.150&   1.000&   1.50&   0.5&   3.7&   6.1&   6.1& SOAR\\
 20552& B&    5.5&   66.0483& $-$57.0720& 2272&   0.054&   0.140&   0.900&   9.0 &   0.0&   3.9&   7.9&   7.9& NICI\\
 20598& B&    3.2&   66.1747& $-$8.7524&  770&   0.029&   0.150&   1.000&   1.50&   0.5&   3.2&   3.7&   3.7& SOAR\\
 21963& B&    8.2&   70.8207& $-$9.6182& 2272&   0.054&   0.140&   0.900&   9.0 &   0.0&   3.9&   7.9&   7.9& NICI\\
 22611& B&   99.6&   72.9498& $-$34.2214&  770&   0.042&   0.150&   1.000&   1.50&   0.5&   3.4&   3.9&   3.9& SOAR
\enddata
\tablecomments{Bouy2008: \citet{Bouy2008}; 
Burg2005: \citet{Burg2005};
Clo2003: \citet{Clo2003};  
Cvn2010: \citet{Cvn2010}; 
Jay2001: \citet{Jay2001}; 
Egg2007: \citet{Egg2007};
MH09: \citet{MH09};  
Mug2009: \citet{Mug2009}; 
NICI: \citet{THH10}; 
SAM, SOAR: this work;
Tok2006: \citet{Tok2006};
WDS: Visual micrometer resolution.
} 
\end{deluxetable*}

\begin{deluxetable*}{ l l cc ccc l   }           
\tabletypesize{\scriptsize}         
\tablecaption{Secondary subsystems 
\label{tab:sec} }                    
\tablewidth{0pt}  
\tablehead{
 \colhead{Name} &  
\colhead{Type} &
\colhead{$\rho$}  &
\colhead{$\Delta m$} & 
\colhead{ $\log P$} &
\colhead {${\cal M}_1$} & 
\colhead {${\cal M}_2$} & 
\colhead{Notes} \\
   & &
(arcsec) &
(mag) & 
(days) &
(${\cal M}_\odot$) &   
(${\cal M}_\odot$) &   
}
\startdata 
14954B & S1,V & 0.064  & \ldots& 2.87 & 0.52 & 0.06 & \citet{Mug2009}, false exo-planet  \\  
34065C & S1,v,a & 0.227&4.4~I  & 3.23 & 0.71 & 0.16 & SOAR,  HIP~34052, \citet{Sahlman2011}: $P=4.62$\,yr \\
35554B & S1 & \ldots   & \ldots& 2.09 & 1.22 & 1.12 & HD~57853, \citet{Saar1990}:  triple, 122\,d and $\sim$10\,d  \\
36165B & s  &  \ldots  & \ldots& \ldots & 0.91 & \ldots & HIP~36160. \citet{N04}: RV variable \\
36395C & v,S1 &  0.090 &1.2~I  & 3.33 & 0.56 & 0.37  & SOAR,  NLTT~17952 (F -- unresolved) \\
36485D&  v,S1 &  0.114 &2.5~I  & 3.43 & 1.05 & 0.61  & SOAR, HIP~36497, \citet{Halb2012}: $P=7.41$\,yr  \\
38908BC& v    &  2.3      & 3.6~V & 4.96 & 0.60 & 0.21 & JSP~208BC \\
43557B & s2   & \ldots & \ldots   & 1.0? & 0.37 & 0.37 &  \citet{Fuhrmann2005}: double-lined\\ 
45170E & v    & 0.53   &  0    & 4.61 & 0.05 & 0.05    & GJ~337C, \citet{Burg2005}: L8/T brown dwarfs  \\
45734B & s2   & \ldots &\ldots &\ldots& 0.93 &0.74    &   \citet{Desidera2006}: double-lined \\
46535A & v    & 0.70   & 3.1~V & 4.62 & 1.21 & 0.76 & HDS~1360, B=HIP~46253 is primary \\
49520B & v    & 0.213  & 3.6~K   & 4.16 & 0.99 & 0.26 & \citet{THH10} \\
53172B & v    &  0.18  &  0    & 4.32 & 0.09 & 0.09 & SAM \\
59021B & S2   &\ldots  & \ldots& 2.17 & 0.84 & 0.67 & double-lined (D.~Latham, private communication)  \\
66676BC& v &   0.918   &3.7~I  & 5.05 & 0.96 & 0.38 & SOAR  \tablenotemark{a} \\
72235B & v  &   0.412  &4.1~I  & 4.44 & 0.70 & 0.17 & SOAR \\
76435C & v  &   0.056  &0.4~I  & 3.15 & 0.70 & 0.66 & SOAR \\
79980B & v  &   0.040  &1.4~I  & 2.70 & 1.16 & 1.00 & SOAR,  HIP~79979, \citet{N04}: RV variable  \\
85342B & v,s,a &1.01   &2.4~I  & 5.09 & 0.93 & 0.63 & SOAR \tablenotemark{b}  \\ 
101806B&  v & 0.42     & 0.79~K& 4.72 & 0.26 & 0.19 & \citet{Egg2007}, A is exo-host \\
113579B&  v &  1.84    & 0.55~V & 5.13 & 0.70 & 0.65 & RST~1154, HIP~113597 \\
114702B&  v & 0.053    & 1.2~K  & 2.91 & 0.90 & 0.66 & \citet{Tok2006}, triple \\
116106BC& v & 0.586    & 2.4~K  & 4.78 & 0.10 & 0.03 & \citet{Bouy2008} 
\enddata
\tablenotetext{a}{HIP~66676B = HD~118735 is resolved here into a triple system: Ca,Cb has a separation of
0\farcs16. The density of background stars is high, so there is still a
chance that B and C are unrelated despite their small separation of 0\farcs92. }
\tablenotetext{b}{HIP~85342B = HIP~85326 has variable RV and {\it Hipparcos} 
acceleration. The new 1\arcsec~ companion  is too
distant to cause RV changes, so B can be triple.}
\end{deluxetable*}


\section{Statistical analysis}
\label{sec:stat}

The  methods  of  statistical  analysis of  secondary  subsystems  are
identical  to those  used  for  the whole  FG-67  sample (see  FG67b).
Separation and  magnitude difference $(\rho, \Delta  m)$ are converted
to period  $P$ (assuming that  the separation equals  semi-major axis)
and  mass ratio $q$  (standard relations  for main  sequence translate
absolute magnitudes  to mass). The detection limits  in $(\rho, \Delta
m)$ are translated to the $(P,q)$ space with the same assumptions.  In
the cases when other detection techniques such as radial velocity (RV)
or {\em  Hipparcos} acceleration are  available, they are  included as
well. Upper  limits on the  periods of potential subsystem  imposed by
the  dynamical stability  are modeled  statistically, as  explained in
FG67b.

\begin{figure}
\epsscale{1.0}
\plotone{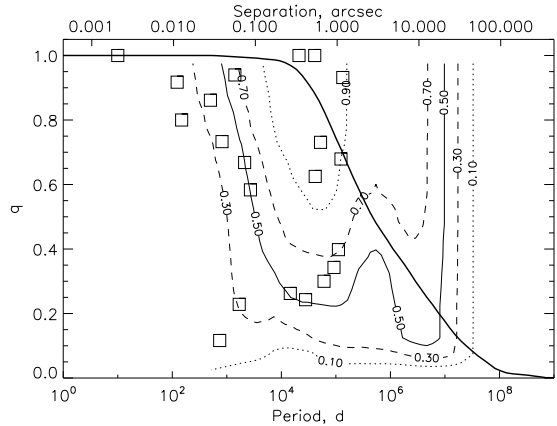}
\caption{Period and mass ratio  of secondary subsystems (squares). The
  contours  show  average   detection  probability.   The  estimated
  fraction  of  dynamically  stable  subsystems  $F_{\rm  dyn}(P)$  is
  plotted  in  thick  line.  The  upper axis  corresponds  to  angular
  separation at a distance of 50\,pc.
\label{fig:pq} 
}
\end{figure}

Figure~\ref{fig:pq} shows the  distribution of secondary subsystems in
the  $(P,q)$  plane  and  the  detection  limits  for  our  sample  of
secondaries. Of the 23 secondary  subsystems, 21 have estimates of $P$
and $q$, the remaining two are spectroscopic binaries with yet unknown
periods.  Subsystems  with long periods  and wide separations  are not
expected because they would be  dynamically unstable in the wide outer
binaries.  The thick line in Figure~\ref{fig:pq} shows the fraction of
dynamically  stable   subsystems  $F_{\rm  dyn}(P)$.    All  secondary
subsystems have  periods shorter than  $10^{5.1}$\,d and sub-arcsecond
separations (the widest pair HIP~113579B = RST~1154 would have $\rho =
1\farcs15$ at  a standard distance  of 50\,pc).  Some of  the surveyed
binaries are quite wide  and could contain subsystems with separations
of  a  few arcseconds.   Such  subsystems  exist  around some  primary
targets, but none were found  around the secondaries, despite the ease
of  their detection.   However, as  noted in  FG67a, there  is  a bias
against  discovery of  wide secondary  components that  are themselves
partially resolved binaries.

Subsystems  with  periods from  $10^3$  to  $10^5$  days have  a  good
detection probability  and are not strongly affected  by the dynamical
stability. There are  12 secondary subsystems with such  periods, or a
raw subsystem frequency of  $12/95 =0.126$.  We correct for incomplete
detection  by  assuming that  the  mass  ratio  $q$ in  the  secondary
subsystems is  distributed as  $q^\beta$ with $\beta=0$  or $\beta=1$.
These two assumptions correspond  to the detection probability of 0.60
and 0.74  and lead to  the detection-corrected companion  frequency of
$0.21\pm0.06$ and  $0.17\pm0.05$, respectively, in  the selected range
of periods. The  uncertainty related to the choice  of $\beta$ is less
than the statistical errors.

\begin{figure}
\epsscale{1.0}
\plotone{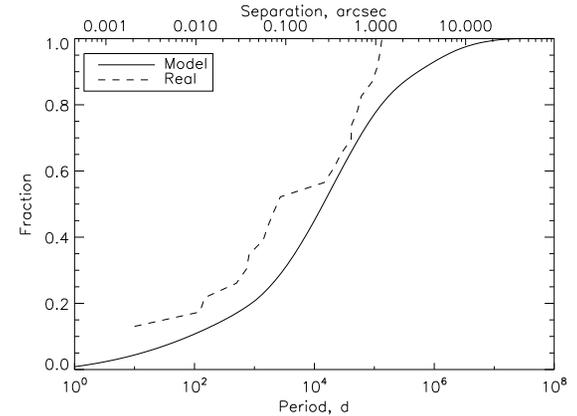}
\caption{Cumulative distribution of periods in the secondary
  subsystems (dashed line) and its model (full line). 
\label{fig:cum} 
}
\end{figure}

By looking  at Figure~\ref{fig:pq},  we note that  the mass  ratios of
secondary subsystems do  not show concentration to $q  \sim 1$ but are
distributed over the  whole range. Fitting the data  by the log-normal
period  distribution  and  the  power-law $q$-distribution  using  the
maximum likelihood method (see FG67b) leads to $\beta = 0.09 \pm 0.04$
for  this sample,  while  $\beta  \sim 1$  was  derived for  secondary
subsystems in  the full  FG-67 sample. So,  we retain the  estimate of
subsystem frequency of $0.21\pm0.06$ as the most plausible.

Binaries  with solar-type  primaries are  described by  the log-normal
period  distribution with  a median  of $10^{4.54}$  days, logarithmic
dispersion  of 2.40,  and  companion fraction  of  0.56 (FG67b).   The
companion fraction in the selected two decades of period is then 0.18,
matching  within errors  the  fraction of  secondary subsystems  found
here.

We tried to evaluate whether  the absence of relatively wide secondary
subsystems  is  significant.   The  statistical model  of  independent
multiplicity developed  in FG67b describes the  period distribution of
subsystems  by   a  product   of  the  log-normal   generating  period
distribution and the dynamical  constraint $F_{\rm dyn}(P)$ imposed by
the  outer binaries.   The full  line in  Figure~\ref{fig:cum} depicts
this model  for the sample  of secondaries studied here.   It predicts
that about a quarter of secondary subsystems (5 out of 23) should have
separations above  1 arcsec.  The  cumulative histogram of  the actual
secondary periods is plotted in  dashed line.  It is not corrected for
the  observational selection  (e.g.  detection  probability  $<0.3$ at
$P<10^3$\,d, see  Figure~\ref{fig:pq}).  Addition of  short undetected
periods would move the dashed  line further to the left.  The emerging
tentative conclusion  is that the periods of  secondary subsystems are
indeed shorter  than predicted  by the dynamical  stability constraint
alone.

\section{Discussion}
\label{sec:disc}

This study confirms that the occurrence of subsystems in the secondary
components of wide binaries is as likely as the binarity of their main
solar-type  primaries. 


With SAM,  we probed binarity of  15 very low  mass (VLM) secondaries.
Their  median mass is  only 0.16\,${\cal  M}_\odot$, the  most massive
(HIP~64056B)  has a mass  of 0.28\,${\cal  M}_\odot$ and  the smallest
(HIP~41211C) is  only 0.11\,${\cal  M}_\odot$. The new  pair HIP~53172
Ba,Bb has  components of 0.10\,${\cal M}_\odot$.   One detected binary
out of  15 means  a 0.07  fraction of subsystems,  which is  less than
$23/95=0.24$ for the whole sample.   However, the resolution of SAM is
inferior in  comparison to  speckle interferometry and  full AO,  so a
lower detection rate is expected.  VLM binaries tend to be closer than
15\,AU,  with a  peak separation  around 5\,AU  \citep{Clo2003}.  This
translates   to  angular   separations  of   0\farcs3   and  0\farcs1,
respectively, at 50\,pc distance.  Most VLM binaries within 67\,pc are
too  close to  be  resolved by  SAM.  If low-mass  components of  wide
binaries are  similar to  other VLM stars,  the detection of  just one
subsystem here is normal.

The  binarity of VLM  stars is  better studied  at diffraction-limited
resolution using large telescopes,  AO, and infrared detectors.  Among
the  23 secondary  subsystems in  Table~\ref{tab:sec},  two (HIP~45170
Ea,Eb and  HIP~116106 BC)  are such VLM  pairs discovered  by adequate
techniques.  Although another  VLM binary was found here  with SAM, we
could not measure accurately its relative position.

On the other  hand, bright secondary components of  larger mass can be
effectively surveyed  by speckle  interferometry which does  reach the
diffraction-limited resolution.  Seven new  pairs (one of them actually
triple) are  added by this  work.  Many secondaries are  bright enough
for RV  monitoring.  It is  obvious from Figure~\ref{fig:pq}  that the
census of secondary subsystems is very poor at short periods, so their
RV survey is needed.

The fraction of secondary subsystems with $x  = \log P$ from 3 to 5 is
found  here to  be $0.21\pm0.06$  or $0.17\pm0.05$,  depending  on the
assumption  about the mass-ratio  distribution.  Adopting  $\beta =1$,
\citet{RoboAO}  found the detection-corrected  frequency of  $0.13 \pm
0.03$ in  a smaller  period range  from 3.5 to  5; this  translates to
$0.17 \pm  0.04$ when  scaled to the  2-dex period interval.   The two
independent samples of wide  secondary components studied by different
instruments gave the same result, enhancing its confidence. 

\citet{RoboAO}  found  that  the  presence of  secondary  and  primary
subsystems  is  correlated.  The  southern  sample  of 95  secondaries
studied  here contains  36 primary  subsystems, 8  of which  also have
secondary subsystems (i.e. are  2+2 quadruples). If the occurrence and
discovery   of   primary  and   secondary   subsystems  are   mutually
independent, the expected number of coincidences is $95 \times (23/95)
\times  (36/95)  = 8.7$.   Therefore,  this  smaller  sample shows  no
evidence of the correlation discovered by \citet{RoboAO} and confirmed
in FG67b.

By  gathering data on  hierarchical multiple  systems, we  advance the
understanding of their origin.  To first order, hierarchical multiples
can be described  by picking randomly inner and  outer subsystems from
the same  generating distribution of  periods and keeping  only stable
configurations.   However, the reality  deviates from  this simplistic
model of independent multiplicity in  several ways.  Here we noted the
lack of  relatively wide secondary subsystems (if  present, they would
have been readily detected).

\citet{Clo2003} believe  that short periods of VLM  binaries cannot be
explained  by  their ejection  from  young  clusters.  Similarly,  the
secondary subsystems  are closer than  allowed by the  outer binaries,
meaning that the  dynamics alone is not a  valid explanation for their
short  periods.   Rather,  this   feature  could  be  related  to  the
correlation between  mass and  angular momentum in  the epoch  of mass
accretion when  the properties of  nascent binaries and  multiples are
established.

\acknowledgments The software for reducing SAM images was developed by
L.~Fraga. C.~Brice\~no  advised me on photometry.  This  work used the
SIMBAD   service   operated  by   Centre   des  Donn\'ees   Stellaires
(Strasbourg, France),  bibliographic references from  the Astrophysics
Data System maintained by SAO/NASA, the Washington Double Star Catalog
maintained at  USNO, and data from  the Sloan Digital  Sky Survey DR9.
Funding  for  SDSS-III has  been  provided  by  the Alfred  P.   Sloan
Foundation,  the  Participating  Institutions,  the  National  Science
Foundation, and the  U.S. Department of Energy Office  of Science. 

{\it Facilities:}  \facility{SOAR}.







\end{document}